\documentclass[a4paper]{artikel3}
\pdfoutput=1

\usepackage{lmodern}
\usepackage[round]{natbib}
\newcommand{\citeg}[1]{\citep[e.g.][]{#1}}
\usepackage[sc]{mathpazo}
\usepackage{inconsolata}
\usepackage{amsmath,amssymb}
\usepackage{url}
\usepackage{verbatim}
\usepackage{graphicx}
\usepackage{color}
\definecolor{mypine}{rgb}{0.05,0.45,0.05}
\definecolor{mydarkred}{rgb}{0.75,0.05,0.05}
\definecolor{mydarkblue}{rgb}{0.05,0.05,0.75}

\newcommand{\Blue}{\textcolor{mydarkblue}}
\newcommand{\Green}{\textcolor{mypine}}
\usepackage{microtype}
\usepackage{nicefrac}
\usepackage{booktabs}
\usepackage{listings}

\newcommand{\te}{\!=\!}
\newcommand{\tp}{\!+\!}

\newcommand{\ttimes}{\!\times\!}
\newcommand{\la}{\!\leftarrow\!}

\newcommand{\br}{{\bf r}}
\newcommand{\bc}{{\bf c}}

\newcommand{\pdd}[2]{\frac{\partial #1}{\partial #2}}
\newcommand{\id}{\mathrm{d}}
\DeclareMathOperator{\Tr}{Tr}
\DeclareMathOperator{\mvec}{vec}
\DeclareMathOperator{\vech}{vech}
\DeclareMathOperator{\tril}{tril}
\newcommand{\kron}{\otimes}

\newcommand{\chol}{L}
\newcommand{\dchol}{\mathrm{d}\kern-0.5pt L}
\newcommand{\LLT}{\Sigma}
\newcommand{\A}{A}

\newcommand{\blas}{\texttt{BLAS}}
\newcommand{\lapack}{\texttt{LAPACK}}
\newcommand{\matlab}{\texttt{Matlab}}
\newcommand{\octave}{\texttt{Octave}}
\newcommand{\R}{\texttt{R}}
\newcommand{\python}{\texttt{Python}}
\newcommand{\numpy}{\texttt{NumPy}}

\usepackage[pdfborder={0 0 0}]{hyperref}

\title{\vspace*{-1cm}Differentiation of the Cholesky decomposition\vspace*{-0.5cm}}

\author{Iain Murray}

\date{February 2016}

\begin{document}

\maketitle

\abstract{We review strategies for differentiating matrix-based
computations, and derive symbolic and algorithmic update rules for
differentiating expressions containing the Cholesky decomposition. We
recommend new `blocked' algorithms, based on differentiating the Cholesky
algorithm \texttt{DPOTRF} in the \lapack\ library, which uses `Level~3'
matrix-matrix operations from \blas\@, and so is cache-friendly and easy to
parallelize. For large matrices, the resulting
algorithms are the fastest way to compute Cholesky derivatives, and
are an order of magnitude faster than the algorithms in common usage. In
some computing environments, symbolically-derived updates are faster for
small matrices than those based on differentiating Cholesky algorithms.
The symbolic and algorithmic approaches can be combined to get the best of
both worlds.}

\section{Introduction}

The Cholesky decomposition $\chol$ of a symmetric positive definite matrix
$\LLT$ is the unique lower-triangular matrix with positive diagonal
elements satisfying $\LLT = \chol\chol^\top$\kern-2pt. Alternatively, some
library routines compute the upper-triangular decomposition $U \te
\chol^\top$\kern-2pt.
This note compares ways to differentiate the function $\chol(\LLT)$, and
larger expressions containing the Cholesky decomposition
(\autoref{sec:tasks}). We consider compact symbolic results
(\autoref{sec:symbolic}) and longer algorithms (\autoref{sec:ad}).

Existing computer code that differentiates expressions containing
Cholesky decompositions often uses an algorithmic approach proposed by
\citet{smith1995}. This approach results from manually applying the ideas
behind `automatic differentiation' \citeg{baydin2015} to a numerical
algorithm for the Cholesky decomposition. Experiments by \citet{walter2011}
suggested that\,---\,despite conventional wisdom\,---\,computing symbolically-derived
results is actually faster. However, these experiments were based on
differentiating slow algorithms for the Cholesky decomposition. In this
note we introduce `blocked' algorithms for propagating Cholesky derivatives
(\autoref{sec:ad}), which use cache-friendly and easy-to-parallelize
matrix-matrix operations. In our implementations (\autoref{sec:code}),
these are faster than all previously-proposed methods.

\section{Computational setup and tasks}
\label{sec:tasks}

\textsl{This section can be safely skipped by readers familiar with ``automatic
differentiation'', the $\dot\LLT$ notation for ``forward-mode
sensitivities'', and the $\bar\LLT$ notation for ``reverse-mode
sensitivities'' \citeg{giles2008}.}

We consider a sequence of computations,
\begin{equation}
    x
    \;\rightarrow\; \LLT
    \;\rightarrow\; \chol
    \;\rightarrow\; f,
    \label{eqn:computation}
\end{equation}
that starts with an input $x$, computes an intermediate symmetric
positive-definite matrix $\LLT$, its lower-triangular Cholesky decomposition~$\chol$, and then
a final result $f$. Derivatives of the overall computation
\smash{\hbox{$\pdd{f}{x}$}}, can be decomposed into reusable parts with the
chain rule. However, there are multiple ways to proceed, some much better than
others.

\smallskip
\textbf{Matrix chain rule:~} It's tempting to simply write down the chain
rule for the overall procedure:
\begin{equation}
    \pdd{f}{x} =
    \sum_{i, j\le i}
    \sum_{k, l\le k}
    \pdd{f}{\chol_{ij}}
    \pdd{\chol_{ij}}{\LLT_{kl}}
    \pdd{\LLT_{kl}}{x},
    \label{eqn:chainidx}
\end{equation}
where we only sum over the independent elements of symmetric matrix $\LLT$ and
the occupied lower-triangle of $\chol$.
We can also rewrite the same chain rule in matrix form,
\begin{equation}
    \pdd{f}{x} =
    \pdd{f}{\vech(\chol)}
    \pdd{\vech(\chol)}{\vech(\LLT)}
    \pdd{\vech(\LLT)}{x},
    \label{eqn:chainmat}
\end{equation}
where the $\vech$ operator creates a vector by stacking the lower-triangular
columns of a matrix. A derivative $\pdd{y}{z}$ is a matrix or vector,
with a row for each element of $y$ and a column for each element of $z$,
giving a row vector if $y$ is a scalar, and a column vector if $z$ is a scalar.

The set of all partial derivatives $\left\{\pdd{\chol_{ij}}{\LLT_{kl}}\right\}$, or
equivalently the matrix $\pdd{\vech(\chol)}{\vech(\LLT)}$, contains $O(N^4)$
values for the Cholesky decomposition of an $N\ttimes N$ matrix. Explicitly
computing each of the terms in equations \eqref{eqn:chainidx} or
\eqref{eqn:chainmat} is inefficient, and simply not practical for large
matrices.

We give expressions for these $O(N^4)$ derivatives
at the end of \autoref{sec:symbolic}
for completeness,
and because they might be useful for analytical study. However, the
computational primitives we really need are methods to accumulate the terms in
the chain rule moving left (forwards) or right (backwards), without creating
enormous matrices. We outline these processes now, adopting the `automatic
differentiation' notation used by \citet{giles2008} and others.

\smallskip
\textbf{Forwards-mode accumulation:~}
We start by computing
a matrix of sensitivities for the first stage of the computation, with elements
$\dot{\LLT}_{kl} \te \pdd{\LLT_{kl}}{x}$.
If we applied an infinitesimal perturbation to the input $x \la x \tp \id{x}$,
the intermediate matrix
would be perturbed by
$\id\LLT\te\dot\LLT\id{x}$.
This change
would in turn perturb the output of the Cholesky decomposition
by $\dchol\te\dot\chol\id{x}$,
where $\dot{\chol}_{ij} \te \pdd{\chol_{ij}}{x}$. We would like to compute the
sensitivities of the Cholesky decomposition, $\dot{\chol}$, from
the sensitivities of the input
matrix $\dot{\LLT}$ and other `local' quantities ($L$ and/or $\Sigma$), without needing to consider
where these came from. Finally, we would compute the required result
\smash{\hbox{$\dot{f} \te \pdd{f}{x}$}} from $\chol$ and $\dot{\chol}$, again
without reference to downstream computations (the Cholesky decomposition).

The \emph{forwards-mode} algorithms in this note describe
how to compute the reusable function
$\dot{\chol}(\chol, \dot\LLT)$, which propagates the effect of a
perturbation forwards through the Cholesky decomposition. The computational
cost will have the same scaling with matrix size as the Cholesky decomposition.
However, if we want the derivatives with respect to $D$ different inputs to the
computation, we must perform the whole forwards propagation $D$ times, each time
accumulating sensitivities with respect to a different input~$x$.

\smallskip
\textbf{Reverse-mode accumulation:~} We can instead accumulate derivatives by starting
at the other end of the computation sequence~\eqref{eqn:computation}. The
effect of perturbing the final stage of the computation is
summarized by a matrix with elements $\bar\chol_{ij} \te
\pdd{f}{\chol_{ij}}$. We need to `back-propagate' this summary to
compute the sensitivity of the output with respect to the downstream matrix,
\smash{$\bar\LLT_{kl} \te \pdd{f}{\LLT_{kl}}$}. In turn, this signal
is back-propagated to compute $\bar{x}\te\pdd{f}{x}$, the target of our
computation, equal to $\dot{f}$ in the forwards propagation above.

The \emph{reverse-mode} algorithms in this note describe
how to construct the reusable function
$\bar{\LLT}(\chol, \bar\chol)$, which propagates the effect of a
perturbation in the Cholesky decomposition backwards, to compute the effect
of perturbing the original positive definite matrix.
Like forwards-mode propagation, the computational cost has the same scaling with matrix size as the Cholesky
decomposition. Reverse-mode differentiation or `back-propagation' has the
advantage that $\bar{\LLT}$ can be reused to compute derivatives with
respect to multiple inputs. Indeed if the input $x$ to the sequence of
computations~\eqref{eqn:computation} is a $D$-dimensional vector, the
cost to obtain all $D$ partial derivatives $\nabla_x f$ scales
the same as a single forwards computation of~$f$. For $D$-dimensional
inputs, reverse-mode differentiation scales a factor of $D$ times better
than forwards-mode.

Reverse-mode computations can have greater memory requirements than forwards
mode, and are less appealing than forwards-mode if there are more outputs of the
computation than inputs.

\section{Symbolic differentiation}
\label{sec:symbolic}

It is not immediately obvious whether a small, neat symbolic form should
exist for the derivatives of some function of a matrix, or whether the
forward- and reverse-mode updates are simple to express. For the Cholesky
decomposition, the literature primarily advises using algorithmic update
rules, derived from the algorithms for numerically evaluating the original
function \citep{smith1995,giles2008}.
However, there are also fairly small algebraic expressions for the
derivatives of the Cholesky decomposition, and for forwards- and
reverse-mode updates.

\textbf{Forwards-mode:~}
\citet{sarkka2013} provides a short derivation of a forwards propagation
rule (his Theorem~A.1), which we adapt to the notation used here.

An infinitesimal perturbation to the expression $\LLT = \chol\chol^\top$ gives:
\begin{equation}
    \id\LLT = \dchol\kern0.5pt\chol^\top + \chol\kern0.5pt\dchol^\top.
\end{equation}
We wish to re-arrange to get an expression for $\dchol$. The trick is to
left-multiply by $\chol^{-1}$ and right-multiply by $\chol^{-\top}$:
\begin{equation}
    \chol^{-1}\id\LLT\kern1pt\chol^{-\top} = \chol^{-1}\dchol + \kern0.5pt\dchol^\top\chol^{-\top}.
\end{equation}
The first term on the right-hand side is now lower-triangular. The second term
is the transpose of the first, meaning it is upper-triangular and has the same
diagonal. We can therefore remove the second term by applying a function $\Phi$
to both sides, where $\Phi$ takes the lower-triangular part of a matrix and
halves its diagonal:
\begin{equation}
    \Phi(\chol^{-1}\id\LLT\kern1pt\chol^{-\top}) = \chol^{-1}\dchol, \qquad
    \text{where~}\; \Phi_{ij}(A) =
    \begin{cases}
        A_{ij} & i > j\\
        \tfrac{1}{2}A_{ii} & i = j\\
        0 & i < j.\\
    \end{cases}
    \label{eqn:phi}
\end{equation}
Multiplying both sides by $\chol$ gives us the perturbation of the
Cholesky decomposition:
\begin{equation}
    \dchol = \chol\kern1pt\Phi(\chol^{-1}\id\LLT\kern1pt\chol^{-\top}).
    \label{eqn:dchol}
\end{equation}
Substituting the forward-mode sensitivity relationships
$\id\LLT\te\dot\LLT\id{x}$ and $\dchol\te\dot\chol\id{x}$ (\autoref{sec:tasks}),
immediately gives a forwards-mode update rule, which is easy to implement:
\begin{equation}
    \fbox{$\displaystyle \dot{\chol} = \chol\kern1pt\Phi(\chol^{-1}\dot\LLT\kern1pt\chol^{-\top})$.}
    \label{eqn:symfwd}
\end{equation}
The input perturbation $\dot\LLT$ must be a symmetric matrix, $\dot\LLT_{kl} =
\dot\LLT_{lk} = \pdd{\LLT_{kl}}{x}$, because $\LLT$ is assumed to be symmetric
for all inputs~$x$.

\medskip
\textbf{Reverse-mode:~}
We can also obtain a neat symbolic expression for
the reverse mode updates.
We substitute \eqref{eqn:dchol} into $\id{f} \te
\Tr(\bar{\chol}^\top\id{\chol})$, and with a few lines of manipulation,
rearrange it into the form $\id{f} \te \Tr(S^\top\id{\LLT})$.
\citet{brewer1977}'s Theorem~1 then implies that for a symmetric matrix $\LLT$,
the symmetric matrix containing reverse mode sensitivities will be:
\begin{equation}
    \fbox{$
    \bar\LLT = S + S^\top - \mathrm{diag}(S), \qquad
    \text{where~\;} S = \chol^{-\top}\Phi(\chol^\top\bar\chol)\chol^{-1},
    $}
    \label{eqn:symrev}
\end{equation}
where $\mathrm{diag(S)}$ is a diagonal matrix containing the diagonal elements
of $S$, and function $\Phi$ is still as defined in~\eqref{eqn:phi}.

Alternatively, a lower-triangular matrix containing the independent
elements of $\bar\LLT$ can be constructed as:
\begin{equation}
    \fbox{$\displaystyle
    \tril(\bar\LLT) = \Phi(S + S^\top)
    = \Phi\big(\chol^{-\top}(P + P^\top)\chol^{-1}\big), \qquad \text{where~\;} P = \Phi(\chol^\top\bar\chol)
    ,
    $}
    \label{eqn:symrevtril}
\end{equation}
with $S$ as in \eqref{eqn:symrev}, and using function $\Phi$ again from \eqref{eqn:phi}.

Since first writing this section we have discovered two similar reverse-mode
expressions \citep{walter2011,koerber2015}. It seems likely that other authors
have also independently derived equivalent results, although these update
rules do not appear to have seen wide-spread use.

\medskip
\textbf{Matrix of derivatives:~}
By choosing the input of interest to be $x\te\LLT_{kl}\te\LLT_{lk}$, and fixing the
other elements of $\LLT$, the sensitivity $\dot\LLT$ becomes a matrix of
zeros except for ones at $\dot\LLT_{kl}\te\dot\LLT_{lk}\te1$. Substituting into
\eqref{eqn:symfwd} gives an expression for all of the partial derivatives of the
Cholesky decomposition with respect to any chosen element of the covariance matrix.
Some further manipulation,
expanding matrix products as sums over indices,
gives an explicit expression for any element,
\begin{equation}
    \pdd{\chol_{ij}}{\LLT_{kl}} =
    \bigg(\sum_{m>j} \chol_{im}\chol_{mk}^{-1} + \tfrac{1}{2}\chol_{ij}\chol_{jk}^{-1}\bigg)\chol_{jl}^{-1}
    +
    (1-\delta_{kl})\bigg(\sum_{m>j} \chol_{im}\chol_{ml}^{-1} + \tfrac{1}{2}\chol_{ij}\chol_{jl}^{-1}\bigg)\chol_{jk}^{-1}.
    \label{eqn:delements}
\end{equation}
If we compute every $(i,j,k,l)$ element, each one can be evaluated in
constant time by keeping running totals of the sums in
\eqref{eqn:delements} as we decrement $j$ from $N$ to~$1$. Explicitly
computing every partial derivative therefore costs~$\Theta(N^4)$.

These derivatives can be arranged into a matrix, by `vectorizing' the
expression \citep{magnus2007,minka2000b,harmeling2013}. We use a well-known identity involving the $\mvec$ operator, which
stacks the columns of a matrix into a vector, and the Kronecker product $\kron$:
\begin{equation}
    \mvec(ABC) = (C^\top\kern-1pt\kron A)\mvec(B).
    \label{eqn:vec2kron}
\end{equation}
Applying this identity to \eqref{eqn:dchol} yields:
\begin{equation}
    \mvec(\dchol) = (I\kron L)\mvec\left(\Phi\big(\chol^{-1}\id\LLT\chol^{-\top}\big)\right).
\end{equation}
We can remove the function $\Phi$, by introducing a diagonal matrix $Z$ defined such
that $Z\mvec(A) = \mvec\Phi(A)$ for any $N\ttimes N$ matrix $A$. Applying
\eqref{eqn:vec2kron} again gives:
\begin{equation}
    \mvec(\dchol) = (I\kron L)Z(\chol^{-1}\kern-1pt\kron\chol^{-1})\mvec(\id\LLT).
\end{equation}
Using the standard
\emph{elimination matrix} $\mathcal{L}$, and \emph{duplication matrix} $D$
\citep{magnus1980}, we
can convert between the $\mvec$ and $\vech$ of a matrix, where $\vech(A)$
is a vector made by stacking the columns of the lower triangle of $A$.
\begin{equation}
    \vech(\dchol) = \mathcal{L}(I\kron L)Z(\chol^{-1}\kern-1pt\kron\chol^{-1})D\vech(\id\LLT)
    \quad\Rightarrow\quad
    \fbox{$\displaystyle
        \pdd{\vech\chol}{\vech\LLT} = \mathcal{L}(I\kron L)Z(\chol^{-1}\kern-1pt\kron\chol^{-1})D.
    $}
    \label{eqn:bletch}
\end{equation}
This compact-looking result was stated on
\emph{MathOverflow}\footnote{\url{http://mathoverflow.net/questions/150427/the-derivative-of-the-cholesky-factor\#comment450752_167719} --- comment from 2014-09-01}
by pseudonymous user `pete'. It may be useful for further analytical study, but
doesn't immediately help with scalable computation.

\section{Differentiating Cholesky algorithms}
\label{sec:ad}

We have seen that it is inefficient to compute each term in the chain rule,
\eqref{eqn:chainidx} or~\eqref{eqn:chainmat}, applied to a high-level
matrix computation. For Cholesky derivatives the cost is $\Theta(N^4)$,
compared to $O(N^3)$ for the forward- or reverse-mode updates in
\eqref{eqn:symfwd}, \eqref{eqn:symrev}, or~\eqref{eqn:symrevtril}. However,
evaluating the terms of the chain rule applied to any \emph{low-level}
computation\,---\,expressed as a series of elementary scalar
operations\,---\,gives derivatives with the same computational complexity
as the original function \citeg{baydin2015}. Therefore $O(N^3)$ algorithms
for the dense Cholesky decomposition can be mechanically converted into
$O(N^3)$ forward- and reverse-mode update algorithms, which is called
`automatic differentiation'.

\citet{smith1995} proposed taking this automatic differentiation approach,
although presented hand-derived propagation algorithms that could be easily
implemented in any programming environment. \citeauthor{smith1995} also
reported applications to sparse matrices, where automatic differentiation
inherits the improved complexity of computing the Cholesky decomposition.
However, the algorithms that were considered for dense matrices aren't
cache-friendly or easy to parallelize, and will be slow in practice.

Currently-popular numerical packages such as
\numpy, \octave, and \R\ \citep{numpy,octave,R}
compute the Cholesky decomposition using the \lapack\ \hbox{library} \citep{lapack}.
\lapack\ implements \emph{block algorithms} that express computations as
cache-friendly, parallelizable `Level~3 \blas' matrix-matrix operations
that are fast on modern architectures. \citet{dongarra1990} described the
Level~3 \blas\ operations, including an example block implementation of a
Cholesky decomposition. For large matrices, we have sometimes found
\lapack's routine to be $50\!\times$ faster than a C or Fortran
implementation of the Cholesky algorithm considered by \citet{smith1995}.
Precise timings are machine-dependent, however it's clear that any large dense matrix
computations, including derivative computations, should be implemented
using blocked algorithms where possible\footnote{Historical note: It's
entirely reasonable that \citet{smith1995} did not use blocked algorithms.
Primarily, \citeauthor{smith1995}'s applications used sparse computations.
In any case, blocked algorithms weren't universally adopted until later.
For example, \matlab\ didn't incorporate \lapack\ until 2000,
\url{http://www.mathworks.com/company/newsletters/articles/matlab-incorporates-lapack.html}.}.

Block routines, like those in \lapack, ultimately come down to elementary
scalar operations inside calls to \blas\ routines. In principle, automatic
differentiation tools could be applied. However, the source code and
compilation tools for the optimized \blas\ routines for a particular
machine are not always available to users. Even if they were, automatic
differentiation tools would not necessarily create cache-friendly
algorithms. For these reasons \citet{walter2011} used symbolic
approaches (\autoref{sec:symbolic}) to provide update rules based on
standard matrix-matrix operations.

An alternative approach is to extend the set of elementary routines understood
by an automatic differentiation procedure to
the operations supported by \blas\@.
We could then pass derivatives through the
Cholesky routine implemented by \lapack, treating the best available matrix-matrix
routines as black-box functions. \citet{giles2008}
provides an excellent tutorial on deriving forward- and reverse-mode update
rules for elementary matrix operations, which we found invaluable for deriving
the algorithms that follow%
\footnote{Ironically, \citet{giles2008} also
considered differentiating the Cholesky decomposition but, like
\citet{smith1995}, gave slow scalar-based algorithms.}.
While his results can largely be found in materials already mentioned \citep{magnus2007,minka2000b,harmeling2013},
\citeauthor{giles2008} emphasised forwards- and reverse-mode update
rules, rather than huge objects like \eqref{eqn:bletch}.

In the end, we didn't follow an automatic differentiation procedure exactly.
While we derived derivative propagation rules from the structure of the Cholesky
algorithms (unlike \autoref{sec:symbolic}), we still symbolically manipulated
some of the results to make the updates neater and in-place. In principle, a
sophisticated optimizing compiler for automatic differentiation could do the same.

\subsection{Level~2 routines}
\label{sec:unblocked}

\lapack\ also provides `unblocked' routines, which use `Level~2' \blas\ operations
\citep{blas2a,blas2b} like matrix-vector products. Although a step up
from scalar-based algorithms, these are intended for small matrices only,
and as helpers for `Level~3' blocked routines (\autoref{sec:blocked}).

The \lapack\ routine \texttt{DPOTF2} loops over columns of an input matrix $\A$,
replacing the lower-triangular part in-place with its Cholesky decomposition. At
each iteration, the algorithm uses a row vector $\br$, a diagonal element
$d$, a matrix $B$, and a column vector $\bc$ as follows:

\medskip
\begin{minipage}{0.5\linewidth}
    \texttt{\textbf{function level2partition}($\A$, $j$)}\\
    \hspace*{2em}\hphantom{$B$}$\llap{$\br$} \,= \A_{j,\,1:j-1}$\\
    \hspace*{2em}\hphantom{$B$}$\llap{$d$} \,= \A_{j,\,j}$\\
    \hspace*{2em}\hphantom{$B$}$\llap{$B$} \,= \A_{j+1:N,\,1:j-1}$\\
    \hspace*{2em}\hphantom{$B$}$\llap{$\bc$} \,= \A_{j+1:N,\,j}$\\
    \hspace*{2em}\texttt{\textbf{return} $\br$, $d$, $B$, $\bc$}
\end{minipage}
\begin{minipage}{0.375\linewidth}
\scalebox{0.8}{$\displaystyle
\mbox{\large where\; $\A =$} \left(
 \begin{array}{ccccc}
   \ddots \\[7pt]
   \relbar\kern-3pt\relbar \mbox{\large $\br$} \relbar\kern-3pt\relbar & \mbox{\large $d$} & & \\[-4pt]
    & | & \ddots\\[-4pt]
    \smash{\text{\Large $B$}}& \mbox{\large $\bc$} & & \ddots\\[-4pt]
    & | & & & \ddots
 \end{array}
\right)
$}
\end{minipage}
\medskip

Here `=' creates a \emph{view} into the matrix $\A$, meaning that in the
algorithm below, `$\leftarrow$' assigns results into the corresponding part of
matrix~$\A$.

\medskip
\begin{minipage}{\linewidth}
\texttt{\textbf{function chol\_unblocked}($\A$)}\\
\hspace*{2em}\textsl{\# If at input $\tril(\A)\te\tril(\LLT)\te\tril(\chol\chol^\top)$\kern-1pt, at output $\tril(\A)\te\chol$.}\\
\hspace*{2em}\texttt{\textbf{for} $j = 1$ \textbf{to} $N$:}\\
\hspace*{2em}\hspace*{2em}\texttt{$\br$, $d$, $B$, $\bc \,=$ level2partition($\A$, $j$)}\\
\hspace*{2em}\hspace*{2em}$d \leftarrow \sqrt{d - \br\br^\top}$\\
\hspace*{2em}\hspace*{2em}$\bc \leftarrow (\bc - B\br^\top) / d$\\
\hspace*{2em}\texttt{\textbf{return} $\A$}
\end{minipage}
\medskip

The algorithm only inspects and updates the lower-triangular part of the
matrix. If the upper-triangular part did not start out filled with zeros,
then the user will need to zero out the upper triangle of the final array
with the $\tril$ function:
\begin{equation}
    \tril(A)_{ij} = \begin{cases}
        A_{ij} & i\ge j\\
        0 & \text{otherwise.}
    \end{cases}
    \label{eqn:tril}
\end{equation}
In each iteration, $\br$ and $B$ are parts of the Cholesky decomposition that
have already been computed, and $d$ and $\bc$ are updated in place, from their
original settings in $\A$ to give another column of the Cholesky decomposition.
The matrix-vector multiplication $B\br^\top$ is a Level~2 \blas\ operation.
These multiplications are the main computational cost of this algorithm.

\smallskip
\textbf{Forwards-mode differentiation:~}

The in-place updates obscure the relationships between parts of the input matrix
and its Cholesky decomposition. We could rewrite the updates more explicitly as
\begin{align}
    \chol_d &= \sqrt{\LLT_d - \chol_\br\chol_\br^\top}\,,\\
    \chol_\bc &= (\LLT_\bc - \chol_B\chol_\br^\top) / \chol_d\,.
\end{align}
Applying infinitesimal perturbations to these equations gives
\begin{align}
    \dchol_d &= \frac{1}{2}(\LLT_d - \chol_\br\chol_\br^\top)^{-1/2}(\id\LLT_d - 2\dchol_\br\chol_\br^\top) \notag \\
             &= \frac{1}{\chol_d}(\id\LLT_d/2 - \dchol_\br\chol_\br^\top)\,, \label{eqn:dub1}\\
    \dchol_\bc &= (\id\LLT_\bc  - \dchol_B\chol_\br^\top - \chol_B\dchol_\br^\top) / \chol_d -  ((\LLT_\bc - \chol_B\chol_\br^\top) / \chol_d^2)\dchol_d \notag \\
               &= (\id\LLT_\bc  - \dchol_B\chol_\br^\top - \chol_B\dchol_\br^\top - \chol_\bc\dchol_d) / \chol_d\,.
               \label{eqn:dub2}
\end{align}
We then get update rules for the forward-mode sensitivities by substituting
their relationships, $\id\LLT\te\dot\LLT\id{x}$ and
$\dchol\te\dot\chol\id{x}$ (\autoref{sec:tasks}), into the equations above.
Mirroring the original algorithm, we can thus convert $\dot\LLT$ to $\dot\chol$
in-place, with the algorithm below:

\medskip
\begin{minipage}{\linewidth}
\texttt{\textbf{function chol\_unblocked\_fwd}($\chol$, $\dot{\A}$)}\\
\hspace*{2em}\textsl{\# If at input $\tril(\dot\A)\te\tril(\dot\LLT)$, at output $\tril(\dot\A)\te\dot\chol$, where $\LLT\te \chol\chol^\top$.}\\
\hspace*{2em}\texttt{\textbf{for} $j = 1$ \textbf{to} $N$:}\\
\hspace*{2em}\hspace*{2em}\texttt{$\br$, $d$, $B$, $\bc \,=$ level2partition($\chol$, $j$)}\\
\hspace*{2em}\hspace*{2em}\texttt{$\dot\br$, $\dot{d}$, $\dot{B}$, $\dot\bc \,=$ level2partition($\dot{\A}$, $j$)}\\
\hspace*{2em}\hspace*{2em}$\dot{d} \leftarrow (\dot{d}/2 - \br\dot\br^\top) / d$\\
\hspace*{2em}\hspace*{2em}$\dot\bc \leftarrow (\dot\bc - \dot{B}\br^\top - B\dot\br^\top - \bc\dot{d}) / d$\\
\hspace*{2em}\texttt{\textbf{return} $\dot{\A}$}
\end{minipage}
\medskip

Alternatively, the Cholesky decomposition and its forward sensitivity can be
accumulated in one loop, by placing the updates from this algorithm after the
corresponding lines in \texttt{chol\_unblocked}.

\smallskip
\textbf{Reverse-mode differentiation:~}

Reverse mode automatic differentiation traverses an algorithm backwards,
reversing the direction of loops and the updates within them. At each step, the
effect $\bar{Z}$ of perturbing an output $Z(A,B,C, \dots)$ is `back-propagated'
to compute the effects $(\bar{A}^{(Z)},\bar{B}^{(Z)},\bar{C}^{(Z)}, \dots)$ of
perturbing the inputs to that step. If the effects of the perturbations are
consistent then
\begin{equation}
    \Tr(\bar{Z}^\top\id Z) =
    \Tr(\bar{A}^{(Z)}{}^\top\id A) +
    \Tr(\bar{B}^{(Z)}{}^\top\id B) +
    \Tr(\bar{C}^{(Z)}{}^\top\id C) + \dots\,,
\end{equation}
and we can find $(\bar{A}^{(Z)},\bar{B}^{(Z)},\bar{C}^{(Z)}, \dots)$ by
comparing coefficients in this equation. If a quantity $A$ is an input to
multiple computations ($X$, $Y$, $Z$, \dots), then we accumulate its total
sensitivity,
\begin{equation}
    \bar{A} = \bar{A}^{(X)} + \bar{A}^{(Y)} + \bar{A}^{(Z)} + \dots,
\end{equation}
summarizing the quantity's effect on the final computation, $\bar{A}_{ij} =
\pdd{f}{A_{ij}}$ (as reviewed in \autoref{sec:tasks}).

Using the standard identities $\Tr(AB)\te\Tr(BA)$,\, $\Tr(A^\top)\te\Tr(A)$,\, and
$(AB)^\top\te B^\top A^\top$, the perturbations from the final line of the
Cholesky algorithm \eqref{eqn:dub2} imply:
\begin{align}
    \Tr(\bar{\chol}_\bc^\top\,\dchol_\bc) =
    \Tr((\bar{\chol}_\bc&/\chol_d)^\top\id\LLT_\bc)  -
    \Tr((\bar{\chol}_\bc\chol_\br/\chol_d)^\top\dchol_B) \notag\\ &-
    \Tr((\bar{\chol}_\bc^\top\chol_B/\chol_d)^\top\dchol_\br) -
    \Tr((\chol_\bc^\top \bar{\chol}_\bc/ \chol_d)^\top\dchol_d)\,.
\end{align}
We thus read off that $\bar{\LLT}_\bc\te\bar{\chol}_\bc/\chol_d$, where the
sensitivities $\bar{\chol}_\bc$ include the direct effect on $f$, provided by
the user of the routine, and the knock-on effects that changing this column
would have on the columns computed to the right. These knock-on effects should
have been accumulated through previous iterations of the reverse propagation
algorithm. From this equation, we can also identify the knock-on effects that
changing $\chol_d$, $\chol_\br$, and $\chol_B$ would have through changing
column $\bc$, which should be added on to their existing sensitivities for
later.

The perturbation \eqref{eqn:dub1} to the other update in the Cholesky algorithm implies:
\begin{equation}
    \Tr(\bar{\chol}_d^\top\dchol_d) =
        \Tr((\bar{\chol}_d/(2\chol_d))^\top\id\LLT_d)
        -\Tr((\bar{\chol}_d\chol_\br/\chol_d)^\top\dchol_\br)\,.
\end{equation}
Comparing coefficients again, we obtain another output of the reverse-mode
algorithm, $\bar{\LLT}_d\te \bar{\chol}_d/(2\chol_d)$. We also add
$\bar{\chol}_d\chol_\br/\chol_d$ to the running total for the sensitivity of
$\chol_\br$ for later updates.

The algorithm below tracks all of these sensitivities, with the updates
rearranged to simplify some expressions and to make an algorithm that can update
the sensitivities in-place.

\medskip
\begin{minipage}{\linewidth}
\texttt{\textbf{function chol\_unblocked\_rev}($\chol$, $\bar\A$)}\\
\hspace*{2em}\textsl{\# If at input $\tril(\bar\A)\te\bar\chol$, at output $\tril(\bar\A)\te\tril(\bar\LLT)$, where $\LLT\te \chol\chol^\top$.}\\
\hspace*{2em}\texttt{\textbf{for} $j = N$ \textbf{to} $1$, \textbf{in steps of} $-1$:}\\
\hspace*{2em}\hspace*{2em}\texttt{$\br$, $d$, $B$, $\bc \,=$ level2partition($\chol$, $j$)}\\
\hspace*{2em}\hspace*{2em}\texttt{$\bar\br$, $\bar{d}$, $\bar{B}$, $\bar\bc \,=$ level2partition($\bar{\A}$, $j$)}\\
\hspace*{2em}\hspace*{2em}\hphantom{$\bar{d}$}$\llap{$\bar{d}$} \leftarrow \bar{d} - \bc^\top\bar\bc/d$\\[3pt]
\hspace*{2em}\hspace*{2em}\hskip-2pt$\bigg[\begin{array}{@{}c@{}}\bar{d}\\[-4pt]\bar{\bc}\\[-0.5pt]\end{array}\bigg] \leftarrow \bigg[\vspace*{-2pt}\begin{array}{@{}c@{}}\bar{d}\\[-4pt]\bar{\bc}\\[-0.5pt]\end{array}\bigg] \big/ d$ \\[1pt]
\hspace*{2em}\hspace*{2em}\hphantom{$\bar{d}$}$\llap{$\bar{\br}$} \leftarrow \bar{\br} - \big[\bar{d}~\,\bar{\bc}^\top\big]\Big[\begin{array}{@{}c@{}}\br\\[-3.5pt]B\end{array}\Big]$\\
\hspace*{2em}\hspace*{2em}\hphantom{$\bar{d}$}$\llap{$\bar{B}$} \leftarrow \bar{B} - \bar{\bc}\br$\\
\hspace*{2em}\hspace*{2em}\hphantom{$\bar{d}$}$\llap{$\bar{d}$} \leftarrow \bar{d}/2$\\
\hspace*{2em}\texttt{\textbf{return} $\bar{\A}$}
\end{minipage}
\medskip

\subsection{Level~3 routines}
\label{sec:blocked}

The \lapack\ routine \texttt{DPOTRF} also updates the lower-triangular part
of an array $\A$ in place with its Cholesky decomposition. However, this
routine updates blocks at a time, rather than single column vectors, using
the following partitions:

\medskip
\begin{minipage}{0.5\linewidth}
    \texttt{\textbf{function level3partition}($\A$, $j$, $k$)}\\
    \hspace*{2em}\hphantom{$D$}$\llap{$R$} \,= \A_{j:k,\,1:j-1}$\\
    \hspace*{2em}\hphantom{$D$}$\llap{$D$} \,= \A_{j:k,\,j:k}$\\
    \hspace*{2em}\hphantom{$D$}$\llap{$B$} \,= \A_{k+1:N,\,1:j-1}$\\
    \hspace*{2em}\hphantom{$D$}$\llap{$C$} \,= \A_{k+1:N,\,j:k}$\\
    \hspace*{2em}\texttt{\textbf{return} $R$, $D$, $B$, $C$}
\end{minipage}
\begin{minipage}{0.375\linewidth}
$\displaystyle
\mbox{where~\,} \A =
\left(
 \begin{array}{ccccc}
   \ddots \\
   R & D & \\
    B & C & \ddots\\
 \end{array}
\right)
$
\end{minipage}
\medskip

Only the lower-triangular part of $D$, the matrix on the diagonal, is
referenced. The algorithm below loops over each diagonal block $D$, updating it
and the matrix $C$ below it. Each diagonal block (except possibly the last) is of size
$N_b\!\times\!N_b$. The optimal block-size $N_b$ depends on the size of the
matrix $N$, and the machine running the code. Implementations of \lapack\
select the block-size with a routine called \texttt{ILAENV}\@.

\medskip
\begin{minipage}{\linewidth}
\texttt{\textbf{function chol\_blocked}($\A$, $N_b$)}\\
\hspace*{2em}\textsl{\# If at input $\tril(\A)\te\tril(\LLT)\te\tril(\chol\chol^\top)$, at output $\tril(\A)\te\chol$, for integer $N_b\!\ge\!1$.}\\
\hspace*{2em}\texttt{\textbf{for} $j = 1$ \textbf{to at most} $N$ \textbf{in steps of} $N_b$:}\\
\hspace*{2em}\hspace*{2em}$k \leftarrow \mbox{\texttt{min}($N$,\, $j\!+\!N_b\!-\!1$)}$\\
\hspace*{2em}\hspace*{2em}\texttt{$R$, $D$, $B$, $C \,=$ level3partition($\A$, $j$, $k$)}\\
\hspace*{2em}\hspace*{2em}$D \leftarrow D - \tril(RR^\top)$\\
\hspace*{2em}\hspace*{2em}$D \leftarrow \mbox{\texttt{chol\_unblocked($D$)}}$\\
\hspace*{2em}\hspace*{2em}$C \leftarrow C - BR^\top$\\
\hspace*{2em}\hspace*{2em}$C \leftarrow C\,\tril(D)^{-\top}$\\
\hspace*{2em}\texttt{\textbf{return} $\A$}
\end{minipage}
\medskip

The computational cost of the blocked algorithm is dominated by Level~3
\blas\ operations for the matrix-matrix multiplies and for solving a
triangular system. The unblocked Level~2 routine from
\autoref{sec:unblocked} (\texttt{DPOTF2} in \lapack) is also called as a
subroutine on a small triangular block. For large matrices it may be worth
replacing this unblocked routine with one that performs more Level~3
operations \citep{gustavson2013}.

\smallskip
\textbf{Forwards-mode differentiation:~}

Following the same strategy as for the unblocked case, we obtained the algorithm below.
As before, the input sensitivities
$\dot\LLT_{ij}\te\pdd{\LLT_{ij}}{x}$ can be updated in-place to give
$\dot\chol_{ij}\te\pdd{\chol_{ij}}{x}$, the sensitivities of the resulting Cholesky
decomposition. Again,
these updates could be accumulated at the same time as
computing the original Cholesky decomposition.

\medskip
\begin{minipage}{\linewidth}
\texttt{\textbf{function chol\_blocked\_fwd}($\chol$, $\dot{\A}$)}\\
\hspace*{2em}\textsl{\# If at input $\tril(\dot\A)\te\tril(\dot\LLT)$, at output $\tril(\dot\A)\te\tril(\dot\chol)$, where $\LLT\te \chol\chol^\top$.}\\
\hspace*{2em}\texttt{\textbf{for} $j = 1$ \textbf{to at most} $N$ \textbf{in steps of} $N_b$:}\\
\hspace*{2em}\hspace*{2em}$k \leftarrow \mbox{\texttt{min}($N$,\, $j\!+\!N_b\!-\!1$)}$\\
\hspace*{2em}\hspace*{2em}\texttt{$R$, $D$, $B$, $C \,=$ level3partition($\chol$, $j$, $k$)}\\
\hspace*{2em}\hspace*{2em}\texttt{$\dot{R}$, $\dot{D}$, $\dot{B}$, $\dot{C} \,=$ level3partition($\dot{\A}$, $j$, $k$)}\\
\hspace*{2em}\hspace*{2em}$\dot{D} \leftarrow \dot{D} - \tril(\dot RR^\top + R\dot R^\top)$\\
\hspace*{2em}\hspace*{2em}$\dot{D} \leftarrow \mbox{\texttt{chol\_unblocked\_fwd($D$, $\dot D$)}}$\\
\hspace*{2em}\hspace*{2em}$\dot{C} \leftarrow \dot{C} - \dot BR^\top - B\dot R^\top$\\
\hspace*{2em}\hspace*{2em}$\dot{C} \leftarrow (\dot C - C\dot D^\top)\,D^{-\top}$\\
\hspace*{2em}\texttt{\textbf{return} $\dot{\A}$}
\end{minipage}
\medskip

The unblocked derivative routine is called as a subroutine. Alternatively,
\texttt{chol\_blocked\_fwd} could call itself recursively with a smaller
block size, we could use the symbolic result \eqref{eqn:symfwd}, or we
could differentiate other algorithms \citeg{gustavson2013}.

Minor detail: The standard \blas\ operations don't provide a routine to
neatly perform the first update for the lower-triangular $\dot{D}$. One
option is to wastefully subtract the full matrix $(\dot RR^\top \tp R\dot
R^\top)$, then zero out the upper-triangle of $\dot D$, meaning that the
upper triangle of $\dot{A}$ can't be used for auxiliary storage.

\smallskip
\textbf{Reverse-mode differentiation:~}

Again, deriving the reverse-mode algorithm and arranging it into a
convenient form was more involved. The strategy is the same as the
unblocked case however, and still relatively mechanical.

\medskip
\begin{minipage}{\linewidth}
\texttt{\textbf{function chol\_blocked\_rev}($\chol$, $\bar{\A}$)}\\
\hspace*{2em}\textsl{\# If at input $\tril(\bar\A)\te\bar\chol$, at output $\tril(\bar\A)\te\tril(\bar\LLT)$, where $\LLT\te \chol\chol^\top$.}\\
\hspace*{2em}\texttt{\textbf{for} $k = N$ \textbf{to no less than} $1$ \textbf{in steps of} $-N_b$:}\\
\hspace*{2em}\hspace*{2em}$j \leftarrow \mbox{\texttt{max}($1$,\, $k\!-\!N_b\!+\!1$)}$\\
\hspace*{2em}\hspace*{2em}\texttt{$R$, $D$, $B$, $C \,=$ level3partition($\chol$, $j$, $k$)}\\
\hspace*{2em}\hspace*{2em}\texttt{$\bar{R}$, $\bar{D}$, $\bar{B}$, $\bar{C} \,=$ level3partition($\bar{\A}$, $j$, $k$)}\\
\hspace*{2em}\hspace*{2em}$ \bar{C} \leftarrow \bar{C}D^{-1}$\\
\hspace*{2em}\hspace*{2em}$ \bar{B} \leftarrow \bar{B} - \bar{C} R$\\
\hspace*{2em}\hspace*{2em}$ \bar{D} \leftarrow \bar{D} - \tril(\bar{C}^\top C)$\\
\hspace*{2em}\hspace*{2em}$ \bar{D} \leftarrow \texttt{chol\_unblocked\_rev}(D, \bar{D})$\\
\hspace*{2em}\hspace*{2em}$ \bar{R} \leftarrow \bar{R} - \bar{C}^\top B - (\bar{D} + \bar{D}^\top)R$\\
\hspace*{2em}\texttt{\textbf{return} $\bar{\A}$}
\end{minipage}
\medskip

The partitioning into columns is arbitrary, so the reverse-mode algorithm
doesn't need to select the same set of blocks as the forwards computation.
Here, when the matrix size $N$ isn't a multiple of the block-size $N_b$,
we've put the smaller blocks at the other edge of the matrix.

As in the blocked forwards-mode update, there is a call to the unblocked
routine, which can be replaced with alternative algorithms. In the
implementation provided (\autoref{sec:code}) we use the
symbolically-derived update~\eqref{eqn:symrevtril}.

\section{Discussion and Future Directions}
\label{sec:discuss}

The matrix operations required by the Cholesky algorithms implemented in
\lapack\ can be implemented with straightforward calls to \blas\@. However,
the forwards- and reverse-mode updates we have derived from these
algorithms give some expressions where only the triangular part of a matrix
product is required. There aren't standard \blas\ routines that implement
exactly what is required, and our implementations must perform unnecessary
computations to exploit the fast libraries available. In future, it would
be desirable to have standard fast matrix libraries that offer a set of
routines that are closed under the rules for deriving derivative updates.

The automatic differentiation tools that have proved popular in machine
learning differentiate high-level array-based code. As a result,
these tools don't have access to the source code of the Cholesky
decomposition, and need to be told how to differentiate it. Theano
\citep{theano1,theano2}, the first tool to be widely-adopted in machine
learning, and AutoGrad \citep{autograd} use the algorithm by
\citet{smith1995}. TensorFlow \citep{abadi2015} in its first release can't
differentiate expressions containing a Cholesky decomposition, but a fork
\citep{gpflow} also uses the algorithm by \citet{smith1995}, as
previously implemented by \citet{gpy}.

The approaches in this note will be an order of magnitude faster for large
matrices than the codes that are in current wide-spread use. Some
illustrative timings are given at the end of the code listing
(\autoref{sec:code}). As the algorithms are only a few lines long, they
could be ported to a variety of settings without introducing any large
dependencies. The simple symbolic expressions (\autoref{sec:symbolic})
could be differentiated using most existing matrix-based tools. Currently
AutoGrad can't repeatedly differentiate the Cholesky decomposition because
of the in-place updates in the \citep{smith1995} algorithm.

The `Level~3' blocked algorithms (\autoref{sec:blocked}) are the fastest
forwards- and reverse-mode update rules for large matrices. However, these
require helper routines to perform the updates on small triangular blocks.
In high-level languages (\matlab, \octave, \python), the `Level~2'
routines\,---\,similar to the algorithms that automatic differentiation
would provide\,---\,are slow, and we recommend using the symbolic updates
(\autoref{sec:symbolic}) for the small matrices instead.

It should be relatively easy to provide similar derivative routines for
many standard matrix functions, starting with the rest of the routines in
\lapack\@. However, it would save a lot of work to have automatic tools to
help make these routines. Although there are a wide-variety of tools for
automatic differentiation, we are unaware of practical tools that can
currently create algorithms as neat and accessible as those made by hand
for this note.

\bibliographystyle{abbrvnat}
\bibliography{bibs}

\newpage
\appendix

\section[Illustrative Python code]{Illustrative \python\ code}
\label{sec:code}

Equations \eqref{eqn:delements} and \eqref{eqn:bletch} were checked
numerically using \octave/\matlab\ code, not provided here.

The rest of the equations and algorithms in this note are illustrated below
using \python\ code that closely follows the equations and pseudo-code.
There are differences due to the note using \matlab/Fortran-style ranges, which are one-based
and inclusive, e.g.\ $1\!:\!3 \te [1,2,3]$. In contrast, \python\ uses
zero-based, half-open ranges, e.g.\ $\mbox{$0\!:\!3$}\!=\,:\!3 \te
[0,1,2]$. The code is also available as \texttt{pseudocode\_port.py} in the
source tar-ball for this paper, available from arXiv.

Development of alternative implementations in multiple programming
languages is on-going. At the time of writing, Fortran code with
\matlab/\octave\ and \python\ bindings, and pure \matlab\ code is available
at \url{https://github.com/imurray/chol-rev}. The Fortran code is mainly
useful for smaller matrices, as for large matrices, the time spent inside
\blas\ routines dominates, regardless of the language used. The code
repository also contains a demonstration of pushing derivatives through a
whole computation (the log-likelihood of the hyperparameters of a Gaussian
process).

\bigskip

\linespread{1.00}

\lstset{
    basicstyle=\fontfamily{zi4}\selectfont\normalsize,
    identifierstyle=,
    stringstyle=\ttfamily,
    keywordstyle=[2],
    showstringspaces=false,
    numbers=left,
    numberstyle=\scriptsize,
    numbersep=2em,
    stringstyle=\Green,
    commentstyle=\small\sl\Blue,
    language=Python,
    morekeywords={with,as}
    }
\begin{lstlisting}[firstnumber=1]
# Demonstration code for Cholesky differentiation
# Iain Murray, February 2016

# These routines need Python >=3.5 and NumPy >= 1.10 for matrix
# multiplication with the infix operator "@". For earlier Python/NumPy,
# replace all uses of "@" with the np.dot() function.

# Tested with Python 3.5.0, NumPy 1.10.4, and SciPy 0.17.0, with MKL
# from Anaconda's distribution with a quad core i5-3470 CPU @ 3.20GHz.

import numpy as np
from numpy import tril
from scipy.linalg import solve_triangular as _solve_triangular

# There are operations that are not performed in place but could be by
# splitting up the operations, and/or using lower-level code. In this
# version of the code, I've instead tried to keep the Python syntax
# close to the illustrative pseudo-code.

# Where the pseudo code contains inverses of triangular matrices, it's
# commonly understood that the matrix product of the inverse with the
# adjacent term should be found by solving the resulting linear system of
# equations. The code below contains some commented-out lines with a
# straightforward rewriting of the pseudocode using inv(), followed by
# equivalent but more efficient lines calling a linear solver (_st()
# defined below). To get an inv function for testing we could do:
#from numpy.linalg import inv
# I'd have to call the underlying LAPACK routines myself to solve these
# systems in-place, as the matrix tranposes haven't worked out to match
# what the SciPy routine can do in-place.

def _st(A, b, trans=0):
    """
    solve triangular system "tril(A) @ x = b", returning x
    
    if trans==1, solve "tril(A).T @ x = b" instead.
    """
    if b.size == 0:
        return b
    else:
        return _solve_triangular(A, b, trans=trans, lower=True)

\end{lstlisting}
\newpage
\begin{lstlisting}[firstnumber=43]
def Phi(A):
    """Return lower-triangle of matrix and halve the diagonal"""
    A = tril(A)
    A[np.diag_indices_from(A)] *= 0.5
    return A

def chol_symbolic_fwd(L, Sigma_dot):
    """
    Forwards-mode differentiation through the Cholesky decomposition
    
    This version uses a "one-line" symbolic expression to return L_dot
    where "_dot" means sensitivities in forwards-mode differentiation,
    and Sigma = L @ L.T.
    """
    # invL = inv(L)
    # return L @ Phi(invL @ Sigma_dot @ invL.T)
    return L @ Phi(_st(L, _st(L, Sigma_dot.T).T))

def chol_symbolic_rev(L, Lbar):
    """
    Reverse-mode differentiation through the Cholesky decomposition
    
    This version uses a short symbolic expression to return
    tril(Sigma_bar) where "_bar" means sensitivities in reverse-mode
    differentiation, and Sigma = L @ L.T.
    """
    P = Phi(L.T @ Lbar)
    #invL = inv(L)
    #return Phi(invL.T @ (P + P.T) @ invL)
    return Phi(_st(L, _st(L, (P + P.T), 1).T, 1))

def level2partition(A, j):
    """Return views into A used by the unblocked algorithms"""
    # diagonal element d is A[j,j]
    # we access [j, j:j+1] to get a view instead of a copy.
    rr = A[j, :j]      # row  
    dd = A[j, j:j+1]   # scalar on diagonal   /      \
    B = A[j+1:, :j]    # Block in corner      | r d  |
    cc = A[j+1:, j]    # column               \ B c  /
    return rr, dd, B, cc

def chol_unblocked(A, inplace=False):
    """
    Cholesky decomposition, mirroring LAPACK's DPOTF2
    
    Intended to illustrate the algorithm only. Use a Cholesky routine
    from numpy or scipy instead.
    """
    if not inplace:
        A = A.copy()
    for j in range(A.shape[0]):
        rr, dd, B, cc = level2partition(A, j)
        dd[:] = np.sqrt(dd - rr@rr)
        cc[:] = (cc - B@rr) / dd
    return A

\end{lstlisting}
\newpage
\begin{lstlisting}[firstnumber=99]
def chol_unblocked_fwd(L, Adot, inplace=False):
    """
    Forwards-mode differentiation through the Cholesky decomposition
    
    Obtain L_dot from Sigma_dot, where "_dot" means sensitivities in
    forwards-mode differentiation, and Sigma = L @ L.T.

    This version uses an unblocked algorithm to update sensitivities
    Adot in place. tril(Adot) should start containing Sigma_dot, and
    will end containing the L_dot. The upper triangular part of Adot
    is untouched, so take tril(Adot) at the end if triu(Adot,1) did
    not start out filled with zeros.

    If inplace=False, a copy of Adot is modified instead of the
    original. The Abar that was modified is returned.
    """
    if not inplace:
        Adot = Adot.copy()
    for j in range(L.shape[0]):
        rr, dd, B, cc = level2partition(L, j)
        rdot, ddot, Bdot, cdot = level2partition(Adot, j)
        ddot[:] = (ddot/2 - rr@rdot) / dd
        cdot[:] = (cdot - Bdot@rr - B@rdot - cc*ddot) / dd
    return Adot

def chol_unblocked_rev(L, Abar, inplace=False):
    """
    Reverse-mode differentiation through the Cholesky decomposition
    
    Obtain tril(Sigma_bar) from L_bar, where "_bar" means sensitivities
    in reverse-mode differentiation, and Sigma = L @ L.T.

    This version uses an unblocked algorithm to update sensitivities
    Abar in place. tril(Abar) should start containing L_bar, and will
    end containing the tril(Sigma_bar). The upper triangular part of
    Adot is untouched, so take tril(Abar) at the end if triu(Abar,1)
    did not start out filled with zeros. Alternatively, (tril(Abar) +
    tril(Abar).T) will give the symmetric, redundant matrix of
    sensitivities.
    
    If inplace=False, a copy of Abar is modified instead of the
    original. The Abar that was modified is returned.
    """
    if not inplace:
        Abar = Abar.copy()
    for j in range(L.shape[0] - 1, -1, -1): # N-1,N-2,...,1,0
        rr, dd, B, cc = level2partition(L, j)
        rbar, dbar, Bbar, cbar = level2partition(Abar, j)
        dbar -= cc @ cbar / dd
        dbar /= dd # / These two lines could be
        cbar /= dd # \ done in one operation
        rbar -= dbar*rr  # / These two lines could be done
        rbar -= cbar @ B # \ with one matrix multiply
        Bbar -= cbar[:,None] @ rr[None,:]
        dbar /= 2
    return Abar

\end{lstlisting}
\newpage
\begin{lstlisting}[firstnumber=156]
def level3partition(A, j, k):
    """Return views into A used by the blocked algorithms"""
    # Top left corner of diagonal block is [j,j]
    # Block size is NB = (k-j)
    R = A[j:k, :j]     # Row block                     /      \
    D = A[j:k, j:k]    # triangular block on Diagonal  |      |
    B = A[k:, :j]      # Big corner block              | R D  |
    C = A[k:, j:k]     # Column block                  \ B C  /
    return R, D, B, C

def chol_blocked(A, NB=256, inplace=False):
    """Cholesky decomposition, mirroring LAPACK's DPOTRF
    
    Intended to illustrate the algorithm only. Use a Cholesky routine
    from numpy or scipy instead."""
    if not inplace:
        A = A.copy()
    for j in range(0, A.shape[0], NB):
        k = min(N, j + NB)
        R, D, B, C = level3partition(A, j, k)
        D -= tril(R @ R.T)
        chol_unblocked(D, inplace=True)
        C -= B @ R.T
        #C[:] = C @ inv(tril(D)).T
        C[:] = _st(D, C.T).T
    return A

def chol_blocked_fwd(L, Adot, NB=256, inplace=False):
    """
    Forwards-mode differentiation through the Cholesky decomposition
    
    Obtain L_dot from Sigma_dot, where "_dot" means sensitivities in
    forwards-mode differentiation, and Sigma = L @ L.T.

    This version uses a blocked algorithm to update sensitivities Adot
    in place. tril(Adot) should start containing Sigma_dot, and will
    end containing the L_dot. Take tril() of the answer if
    triu(Adot,1) did not start out filled with zeros. Unlike the
    unblocked routine, if the upper triangular part of Adot started
    with non-zero values, some of these will be overwritten.

    If inplace=False, a copy of Adot is modified instead of the
    original. The Abar that was modified is returned.
    """
    if not inplace:
        Adot = Adot.copy()
    for j in range(0, L.shape[0], NB):
        k = min(N, j + NB)
        R, D, B, C = level3partition(L, j, k)
        Rdot, Ddot, Bdot, Cdot = level3partition(Adot, j, k)
        Ddot[:] = tril(Ddot) - tril(Rdot @ R.T + R@Rdot.T)
        #chol_unblocked_fwd(D, Ddot, inplace=True) # slow in Python
        Ddot[:] = chol_symbolic_fwd(D, Ddot + tril(Ddot, -1).T)
        Cdot -= (Bdot@R.T + B@Rdot.T)
        #Cdot[:] = (Cdot - C@Ddot.T) @ inv(tril(D)).T
        Cdot[:] = _st(D, Cdot.T - Ddot@C.T).T
    return Adot

\end{lstlisting}
\newpage
\begin{lstlisting}[firstnumber=214]
def chol_blocked_rev(L, Abar, NB=256, inplace=False):
    """
    Reverse-mode differentiation through the Cholesky decomposition
    
    Obtain tril(Sigma_bar) from L_bar, where "_bar" means sensitivities
    in reverse-mode differentiation, and Sigma = L @ L.T.

    This version uses a blocked algorithm to update sensitivities Abar
    in place. tril(Abar) should start containing L_bar, and will end
    containing the tril(Sigma_bar). Take tril(Abar) at the end if
    triu(Abar,1) did not start out filled with zeros. Alternatively,
    (tril(Abar) + tril(Abar).T) will give the symmetric, redundant
    matrix of sensitivities.
    
    Unlike the unblocked routine, if the upper triangular part of Abar
    started with non-zero values, some of these will be overwritten.

    If inplace=False, a copy of Abar is modified instead of the
    original. The Abar that was modified is returned.
    """
    if not inplace:
        Abar = Abar.copy()
    for k in range(L.shape[0], -1, -NB):
        j = max(0, k - NB)
        R, D, B, C = level3partition(L, j, k)
        Rbar, Dbar, Bbar, Cbar = level3partition(Abar, j, k)
        #Cbar[:] = Cbar @ inv(tril(D))
        Cbar[:] = _st(D, Cbar.T, trans=1).T
        Bbar -= Cbar @ R
        Dbar[:] = tril(Dbar) - tril(Cbar.T @ C)
        #chol_unblocked_rev(D, Dbar, inplace=True) # slow in Python
        Dbar[:] = chol_symbolic_rev(D, Dbar)
        Rbar -= (Cbar.T @ B + (Dbar + Dbar.T) @ R)
    return Abar

# Testing code follows

def _trace_dot(A, B):
    """_trace_dot(A, B) = trace(A @ B) = A.ravel() @ B. ravel()"""
    return A.ravel() @ B.ravel()

def _testme(N):
    """Exercise each function using NxN matrices"""
    import scipy as sp
    from time import time
    if N > 1:
        Sigma = np.cov(sp.randn(N, 2*N))
        Sigma_dot = np.cov(sp.randn(N, 2*N))
    elif N == 1:
        Sigma = np.array([[sp.rand()]])
        Sigma_dot = np.array([[sp.rand()]])
    else:
        assert(False)
    tic = time()
    L = np.linalg.cholesky(Sigma)
    toc = time() - tic
    print('Running np.linalg.cholesky:')
    print('   Time taken: %0.4f s' % toc)
    tic = time()
    L_ub = tril(chol_unblocked(Sigma))
    toc = time() - tic
    print('Unblocked chol works: %r'
            % np.all(np.isclose(L, L_ub)))
    print('   Time taken: %0.4f s' % toc)
    tic = time()
    L_bl = tril(chol_blocked(Sigma))
    toc = time() - tic
    print('Blocked chol works: %r'
            % np.all(np.isclose(L, L_bl)))
    print('   Time taken: %0.4f s' % toc)
    tic = time()
    Ldot = chol_symbolic_fwd(L, Sigma_dot)
    toc = time() - tic
    hh = 1e-5
    L2 = np.linalg.cholesky(Sigma + Sigma_dot*hh/2)
    L1 = np.linalg.cholesky(Sigma - Sigma_dot*hh/2)
    Ldot_fd = (L2 - L1) / hh
    print('Symbolic chol_fwd works: %r'
            % np.all(np.isclose(Ldot, Ldot_fd)))
    print('   Time taken: %0.4f s' % toc)
    tic = time()
    Ldot_ub = tril(chol_unblocked_fwd(L, Sigma_dot))
    toc = time() - tic
    print('Unblocked chol_fwd works: %r'
            % np.all(np.isclose(Ldot, Ldot_ub)))
    print('   Time taken: %0.4f s' % toc)
    tic = time()
    Ldot_bl = tril(chol_blocked_fwd(L, Sigma_dot))
    toc = time() - tic
    print('Blocked chol_fwd works: %r'
            % np.all(np.isclose(Ldot, Ldot_bl)))
    print('   Time taken: %0.4f s' % toc)
    Lbar = tril(sp.randn(N, N))
    tic = time()
    Sigma_bar = chol_symbolic_rev(L, Lbar)
    toc = time() - tic
    Delta1 = _trace_dot(Lbar, Ldot)
    Delta2 = _trace_dot(Sigma_bar, Sigma_dot)
    print('Symbolic chol_rev works: %r'
            % np.all(np.isclose(Delta1, Delta2)))
    print('   Time taken: %0.4f s' % toc)
    tic = time()
    Sigma_bar_ub = chol_unblocked_rev(L, Lbar)
    toc = time() - tic
    Delta3 = _trace_dot(Sigma_bar_ub, Sigma_dot)
    print('Unblocked chol_rev works: %r'
            % np.all(np.isclose(Delta1, Delta3)))
    print('   Time taken: %0.4f s' % toc)
    tic = time()
    Sigma_bar_bl = chol_blocked_rev(L, Lbar)
    toc = time() - tic
    Delta4 = _trace_dot(Sigma_bar_bl, Sigma_dot)
    print('Blocked chol_rev works: %r'
            % np.all(np.isclose(Delta1, Delta4)))
    print('   Time taken: %0.4f s' % toc)

if __name__ == '__main__':
    import sys
    if len(sys.argv) > 1:
        N = int(sys.argv[1])
    else:
        N = 500
    _testme(N)

    # Example output for N = 500
    # --------------------------
    # Running np.linalg.cholesky:
    #    Time taken: 0.0036 s
    # Unblocked chol works: True
    #    Time taken: 0.0319 s
    # Blocked chol works: True
    #    Time taken: 0.0356 s
    # Symbolic chol_fwd works: True
    #    Time taken: 0.0143 s
    # Unblocked chol_fwd works: True
    #    Time taken: 0.0592 s
    # Blocked chol_fwd works: True
    #    Time taken: 0.0112 s
    # Symbolic chol_rev works: True
    #    Time taken: 0.0165 s
    # Unblocked chol_rev works: True
    #    Time taken: 0.1069 s
    # Blocked chol_rev works: True
    #    Time taken: 0.0093 s

    # Example output for N = 4000
    # --------------------------
    # Running np.linalg.cholesky:
    #    Time taken: 0.4020 s
    # Unblocked chol works: True
    #    Time taken: 25.8296 s
    # Blocked chol works: True
    #    Time taken: 0.7566 s
    # Symbolic chol_fwd works: True
    #    Time taken: 3.9871 s
    # Unblocked chol_fwd works: True
    #    Time taken: 51.6754 s
    # Blocked chol_fwd works: True
    #    Time taken: 1.2495 s
    # Symbolic chol_rev works: True
    #    Time taken: 4.1324 s
    # Unblocked chol_rev works: True
    #    Time taken: 96.3179 s
    # Blocked chol_rev works: True
    #    Time taken: 1.2938 s

    # Times are machine and configuration dependent. On the same test
    # machine, my Level 3 Fortran implementation is only ~10% faster
    # for N=4000, although can be a lot faster for small matrices.
    # A Level 2 Fortran implementation isn't as bad as the Python
    # version, but is still >15x slower than the blocked Python code.

\end{lstlisting}

\end{document}